*Title:*
DISTINCT DIFFERENCES IN THE NANOSCALE BEHAVIORS OF THE TWIST-BEND LIQUID CRYSTAL PHASE OF A FLEXIBLE LINEAR TRIMER AND HOMOLOGOUS DIMER

*Short title:*
TWIST-BEND PHASE OF A LIQUID CRYSTAL DIMER AND TRIMER

*Authors:* Michael R. Tuchband[1], Daniel A. Paterson[2], Mirosław Salamończyk[3,4], Victoria A. Norman[4], Alyssa N. Scarbrough[5], Ewan Forsyth[2], Edgardo Garcia[6], Cheng Wang[4], John M. D. Storey[2], David M. Walba[5], Samuel Sprunt[3], Antal Jákli[3], Chenhui Zhu[4*], Corrie T. Imrie[2], and Noel A. Clark[1*]

*Author affiliations:*
[1]*Department of Physics and Soft Materials Research Center,*
*University of Colorado, Boulder, CO 80309-0390, USA*

[2]*Department of Chemistry, School of Natural and Computing Sciences,*
*University of Aberdeen, Meston Building, Aberdeen, AB24 3UE, UK*

[3]*Liquid Crystal Institute and Chemical Physics Interdisciplinary Program,*
*Kent State University, Kent, OH 44242, USA*

[4]*Advanced Light Source, Lawrence Berkeley National Laboratory, Berkeley, California 94720, USA*

[5]*Department of Chemistry and Biochemistry and Soft Materials Research Center,*
*University of Colorado, Boulder, CO 80309-0390, USA*

[6]*Laboratório de Química Computacional, Instituto de Química, Universidade de Brasília, Brasília, Brasil*

*\*Corresponding Author information:*
*Michael Tuchband*
michael.tuchband@colorado.edu
*4809801215*
*Department of Physics*
*390 UCB*
*University of Colorado*
*Boulder, CO 80309-0390*

**ABSTRACT**

We synthesized the liquid crystal dimer and trimer members of a series of flexible linear oligomers and characterized their microscopic and nanoscopic properties using resonant soft x-ray scattering and a number of other experimental techniques. On the microscopic scale, the twist-bend phases of the dimer and trimer appear essentially identical. However, while the liquid crystal dimer exhibits a temperature-dependent variation of its twist-bend helical pitch varying from 100 – 170 Å on heating, the trimer exhibits an essentially temperature-independent pitch of 66 Å, significantly shorter than those reported for other twist-bend forming materials in the literature. We attribute this to a specific combination of intrinsic conformational bend of the trimer molecules and a sterically favorable intercalation of the trimers over a commensurate fraction (two-thirds) of the molecular length. We develop a geometric model of the twist-bend phase for these materials with the molecules arranging into helical chain structures, and we fully determine their respective geometric parameters.

**SIGNIFICANCE STATEMENT**

The twist-bend (TB) liquid crystal phase is the newest nematic phase, having only been identified in 2011. Still, there are many outstanding mysteries about the nature of its nanoscale organization and behavior. We elucidate how the number of monomer units in a linear TB oligomer influences the structure of its nanoscale helix, an important structure-property relationship of the TB phase. While a TB dimer exhibits a temperature-dependent variation of its helix pitch, the analogous trimer features a temperature-*independent* helix pitch considerably shorter than that of the dimer and other known TB materials. This study illuminates the scope of possible variations that manifest in the TB phase and represents a substantial step in controlling its nanoscale behavior for technological applications.

**INTRODUCTION**

Liquid crystal (LC) compounds with two or more rigid monomer units connected by flexible linkers can exhibit unique assemblies not encountered in other LCs. LC oligomers bridge the structure-space between conventional small molecule LCs and polymer LCs. This structural motif has generated considerable interest in the LC field because their chemical make-up induces odd-even effects and unique intercalated smectic phases(1, 2).

In addition, there have been a number of accounts of short oligomers which form the twist-bend (TB) LC phase(3–8), a phase which has only recently been identified(9). The TB phase is a nematic (N) characterized by heliconical orientational ordering of the LC director, where the molecules are inclined by a cone angle $\theta_H$ with respect to the helical axis $z$, and precess with azimuthal angle $\varphi(z)$ around $z$, with nanoscale pitch $p_H$, but do so without any accompanying mass density wave in $p_H$(10). The nanoscale TB structure has been investigated in a number of LC systems with a variety of structural motifs, including flexible dimers(10–15), a bent-core mesogen(16), and mixtures of bent LC molecules with other LCs(11, 17–19). However, only one nanoscale structural investigation has previously been performed on TB-forming oligomers of three or more monomeric units(10), although in that case the compound was not a conventional linear trimer but a bent-core hybrid molecule. A number of studies have found through conventional x-ray diffraction techniques that trimers and tetramers tend to be heavily intercalated in the nematic and TB phases(4, 6), but to our knowledge, little other characterization of the nanoscale structure of the TB phase has been carried out on linear unbranched LC $n$-mers with $n > 2$.

Here, we describe the synthesis of an LC molecular trimer which we designate CB6OBO6CB. We perform a number of nanoscale characterization techniques, including resonant soft x-ray scattering (RSoXS), wide-angle x-ray diffraction (WAXS), and freeze-fracture transmission electron microscopy (FFTEM), and contrast the results to those of the analogous dimer molecule CB6OCB which also forms the TB phase(14). While CB6OCB exhibits a TB pitch which varies considerably from $p_{di}$ = 100 – 170 Å, we find that CB6OBO6CB exhibits a pitch $p_{tri}$ = 66 – 67 Å over the ~20°C range of its TB phase, making it effectively temperature-independent and the smallest TB-pitch yet reported. Because this molecule is significantly longer than most other conventional TB-forming LCs, we find such a small pitch extraordinary. We account for this fact by considering the degree of intercalation possible in CB6OBO6CB, which permits a sterically favorable interlocking of molecules over ~2/3 of their molecular length, thereby constraining the flexibility and suppressing fluctuations of the helical structure. Our characterization further permits us to construct detailed geometric models of the TB phase in each of these materials and to make predictions about the behaviors of the analogous higher oligomers.

**RESULTS**

CB6OBO6CB (or the "trimer", for conciseness), contains three rod-like monomer units linked by two flexible spacers (Fig. 1a). The odd number of atoms separating the monomer units ensures oligomer curvature, while the choice of the methylene link between the spacer and the monomeric cyanobiphenyl groups ensures a sufficient bend to make it a promising candidate to form the TB phase, by analogy to CB6OCB (the "dimer")(14).

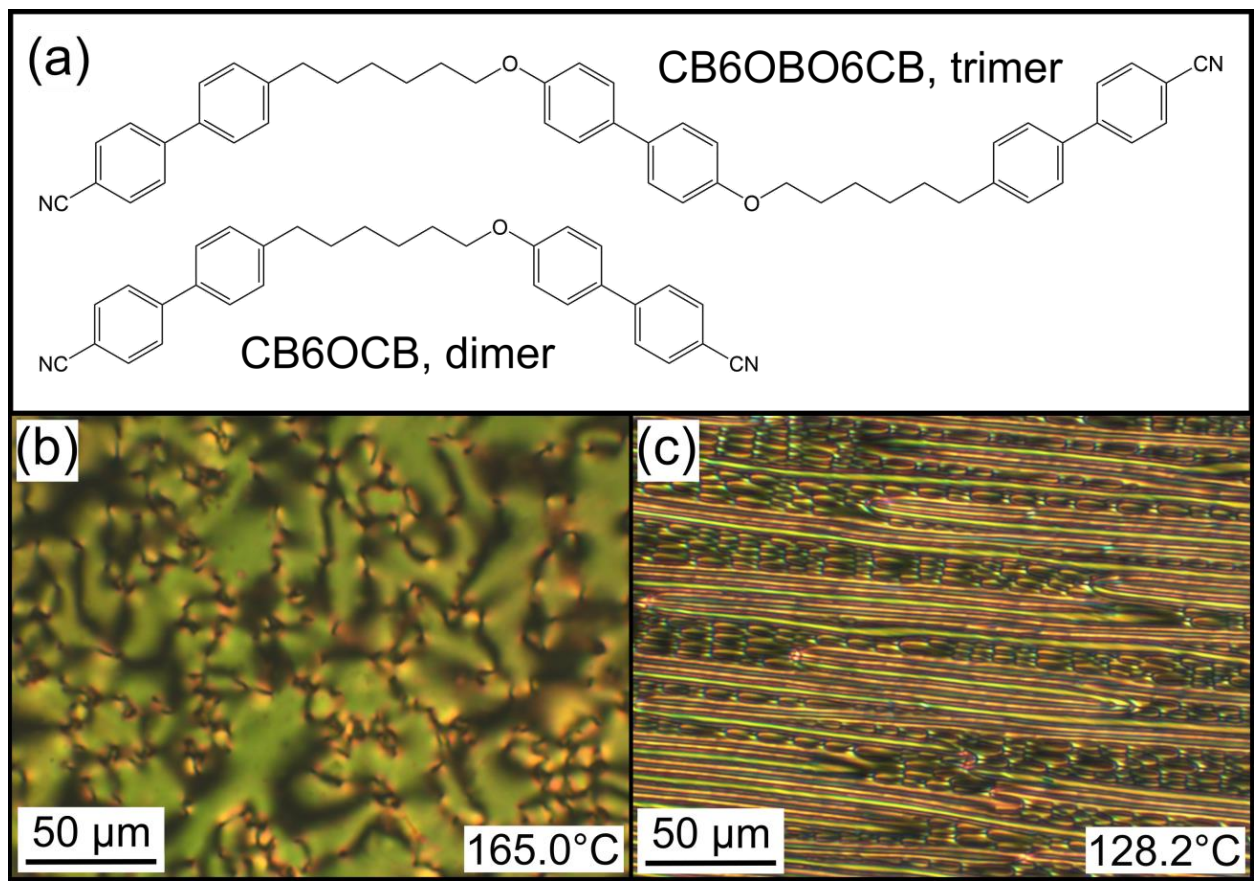

Fig. 1: Molecular structure of CB6OBO6CB and CB6OCB and polarized light microscopy images of the trimer. (a) Molecular structure drawings of CB6OBO6CB and CB6OCB, which we denote "trimer" and "dimer", respectively. Polarized light microscopy images of the trimer in the nematic phase (b) and the TB phase (c).

We characterized the phase behavior of CB6OBO6CB and positively identified the TB phase with a combination of polarized light microscopy (Figs. 1b, c) differential scanning calorimetry (Supplementary Fig. S1), a binary phase diagram of the dimer and trimer (Supplementary Fig. S2), and a contact cell preparation of the trimer with CB7CB(9–11, 13) (Supplementary Fig. S3), a material well-known to exhibit the TB phase. We find the following LC phase behaviors: for CB6OBO6CB, I (isotropic) $\xrightarrow{169°C}$ N (nematic) $\xrightarrow{131°C*}$ TB $\xrightarrow{140°C}$ Cr (crystal) (the * symbol indicates a monotropic phase transition), and for CB6OCB, I $\xrightarrow{155°C}$ N $\xrightarrow{109°C}$ TB $\xrightarrow{99°C}$ Cr.

A typical uniaxial nematic texture forms on cooling from the isotropic phase of the trimer in a unidirectionally-rubbed planar cell (Fig. 1b). On further cooling, the nematic phase transitions into another phase which exhibits a blocky optical texture that then develops into well-defined stripes and focal conic defects (Fig. 1c). These textures persist for ~20°C on steady cooling until the transition into the crystal phase. This behavior is fully consistent with the TB phase observed in LC dimers(9, 20).

The enthalpy of formation ($\Delta H$) and the entropy change ($\Delta S$) associated with the transitions are in the range of expected values for these types of transitions (Fig. 1). It is noteworthy, however, that the entropy change associated with the N – TB transition is greater than would be expected based on the behavior of dimeric materials as well as the temperature width of the preceding nematic phase (Supplementary Fig. S1, refs. (5–7, 21)).

We subsequently performed RSoXS(13) on CB6OCB and CB6OBO6CB by taking 2D detector images on cooling from the nematic phase, where we observed no resonant scattering. We converted these 2D RSoXS detector images into plots of $I(q)$ vs. $q$, the magnitude of the wave vector **q**, by azimuthally averaging the 2D diffractograms about $q = 0$. We then used the Nika x-ray data analysis and processing software to interpolate the 1D plots into a color map with $q$ and the corresponding wavelength $d(q) = 2\pi/q$ plotted as a function of temperature, with the intensity in the 1D plots represented by the color scale in Figs. 2a,b. On cooling from the nematic, CB6OCB develops a peak at $q(T = 108°C) = 0.0375$ Å$^{-1}$, corresponding to the TB helical pitch $d(q) = p_{di}(T = 108°C) \approx 170$ Å (Fig. 2a). The helical pitch rapidly decreases, then begins to saturate near $p_{di}(T = 60°C) \approx 100$ Å before it crystallizes. The determination of a TB pitch of ~90 Å for CB6OCB by FFTEM in ref.(14) roughly accords with our RSoXS measurement, given that FFTEM experiments tend to exhibit the value of the pitch extrapolated to low temperature in the TB phase(19).

CB6OBO6CB, on the other hand, exhibits a very different TB pitch behavior (Fig. 2b). On cooling from the nematic phase, we begin to see unambiguous evidence of scattering from the TB helix at ~130°C. At 124°C, we held the temperature fixed and varied the beamline energy about the carbon K-edge resonance to check that the scattering feature was indeed resonant (Supplementary Fig. S4), confirming that the structure of this phase is modulated in molecular orientation and not related to electron density(13). The scattering signal peaks at $q_{tri}$(high $T$) = 0.094 Å$^{-1}$, corresponding to a helical TB pitch of $p_{tri}$(high $T$) = 67 Å. On further cooling, this peak shifts very gradually to $q_{tri}$(low $T$) = 0.096 Å$^{-1}$, corresponding to a helical pitch of $p_{tri}$(low $T$) = 66 Å. The lack of significant temperature dependence of the pitch on cooling over the ~20°C temperature range is striking, and in stark contrast to other RSoXS measurements performed on TB dimers and their mixtures(13, 15, 18, 19). This TB scattering feature fades away as the sample crystallizes near 110°C. We scanned all accessible RSoXS sample-detector configurations to check for other possible resonant scattering signals from $q$ = 0.0012 – 0.10 Å$^{-1}$ ($d$ = ~60 – 5000 Å) but found none. The value of the pitch obtained from RSoXS is in good agreement to our FFTEM measurement of $p_{tri,FFTEM}$ = 66 Å (Fig. 2c, Supplementary Fig. S5) for the trimer, where we analyzed a number of images to obtain a statistical distribution of the observed periodicities throughout the sample, as described in ref.(19).

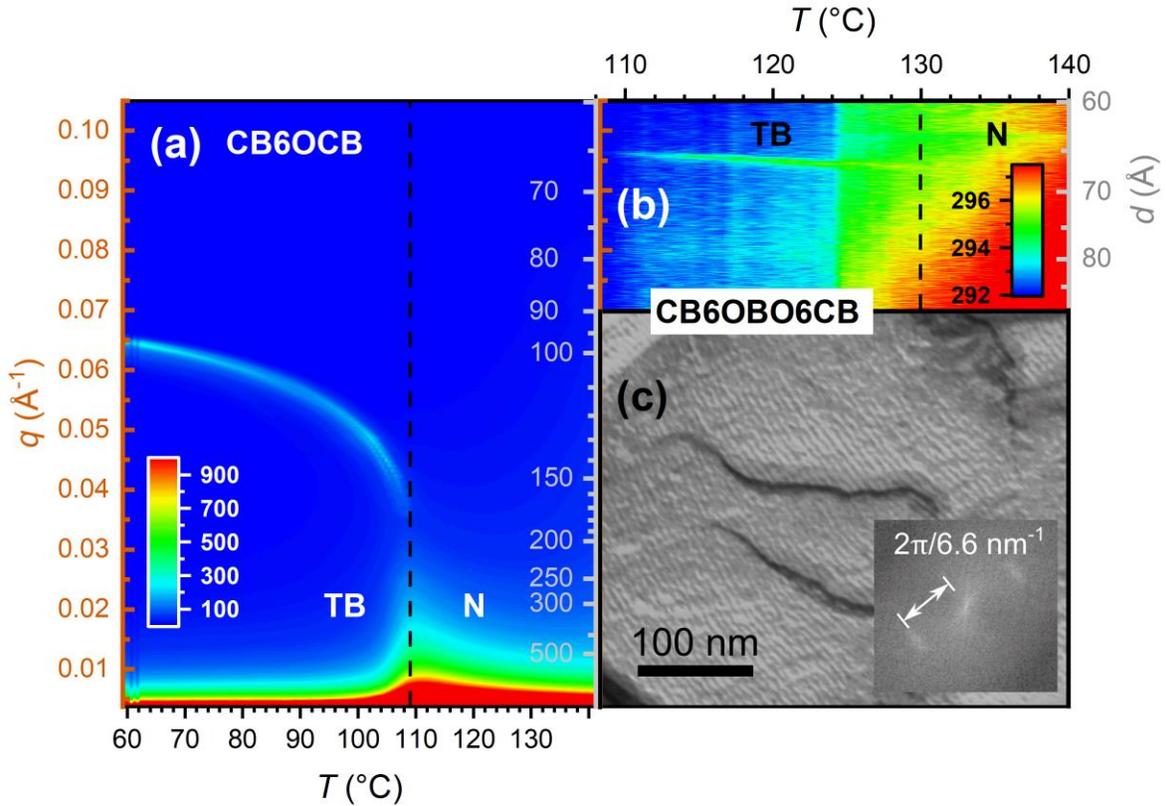

Fig. 2: RSoXS color plots of CB6OCB and CB6OBO6CB with the $q$-axis in orange and the $d$-axis in grey, and an FFTEM image of CB6OBO6CB. (a,b) Color plots are composed of line scans in $q$ of azimuthally averaged 2D detector images from RSoXS expeirments taken as a function of temperature. The $q$ and pitch $d$ scales are the same across both plots. (a) CB6OCB exhibits usual temperature-dependent scattering behavior for conventional TB-forming dimer molecules. In the nematic phase, we observe no scattering features, as expected. On cooling to 108°C, a peak appears near $p_{di}(T = 108°C) \approx$ 170 Å corresponding to the TB helical pitch. On further cooling the helical pitch drops dramatically, then more slowly approaches $p_{di}(T = 60°C) \approx 100$ Å before crystallization. The exposure time was 3 seconds. (b) On cooling CB6OBO6CB from the nematic phase, a scattering feature appears near 130°C corresponding to a helical pitch of $p_{tri} = 67 - 66$ Å, remarkably short compared with that of CB6OCB and other known TB materials. The dashed line denoting the N – TB phase transition temperature was drawn where the TB phase scattering is clearly visible. The plot exhibits a slight discontinuity in the background scattering $T = 124°C$, which we believe to be due to thermal drift while we performed a beamline energy scan. The exposure time was 5 seconds. (c) Representative FFTEM image of CB6OBO6CB in the TB phase exhibiting sinusoidal topographical modulations at $p_{tri,FFTEM} = 6.6$ nm. In different regions of the trimer sample, we observe topographical modulations with a variety of periodicities, but with the most frequently observed periodicity in the sample (when weighted by the area of occurrence) is 6.6 nm (see Supplementary Fig. S5).

Using WAXS, we investigated the diffuse scattering in the trimer corresponding to correlations in mass density along ($q_{z,WAXS}$) and perpendicular to ($q_{\perp,WAXS}$) the TB helix axis. In the TB phase, the scattering feature along the helix axis is present near $q_{z,WAXS}$ = 0.48 Å$^{-1}$ (Supplementary Fig. S6). This corresponds to $d_z$ = 13 Å, or an intercalation of the trimer over nearly 1/3 of the molecular length. This is in agreement with the significant intercalation reported in previous studies of TB and smectic phases of LC linear oligomers(4, 6, 22). The $q_{\perp,WAXS}$ feature corresponds to an average ~5 Å lateral spacing between molecules of CB6OBO6CB.

We carried out birefringence measurements of the dimer and trimer (shown in Supplementary Fig. S7), then used the procedure developed by Meyer *et al.* [28] to calculate an optically-derived cone angle from the birefringence as a function of temperature. The behavior of the optically-derived cone angle in the dimer and trimer is similar to that found for CB7CB. At high temperature in the TB phase, the dimer exhibits a cone angle of ~7°, while the trimer exhibits a cone angle of ~10°. The cone angle saturates near 20° and 17° for the dimer and trimer, respectively, which is somewhat smaller than that found for CB7CB(19, 23).

**DISCUSSION**

The striking differences in the nanoscale behavior of the TB phase in the dimer and trimer must be due to effects from the different number of rod-shaped monomer units in the molecules. A number of review studies have discussed the variations in physical and LC properties as a function of the number of monomer units in small LC oligomers(1, 2, 6, 24, 25), but beyond the work in the LC community, the literature on the properties of linear oligomer homologues is sparse. However, we presently have enough experimental detail to construct a model of the TB organization in each material and to speak to their differences.

The TB nanostructure is determined by a combination of molecular features, including the chemical make-up of the molecules and the relative dimensions of the monomer units, and structural features, including the TB pseudo-layer spacing and helix pitch.

The "pseudo-layer" spacing $s$ in the TB phase is essentially the same feature as that found in aligned nematics, with scattering arcs in $q_z$ elucidating the density modulation along the nematic director $\hat{z}$(26). We can determine $s$ for the TB phase from the $q_{z,\text{WAXS}}$ feature in WAXS experiments for both the dimer(14) and trimer (Supplementary Fig. S6), where $s = d(q_{z,\text{WAXS}}) = 2\pi/q_{z,\text{WAXS}}$. We find the pseudo-layer spacing for the dimer is $s_{\text{di}} = 11.6$ Å(14) and for the trimer is $s_{\text{tri}} = 13$ Å, as denoted in Figs. 3a,b. That $s_{\text{di}}$ and $s_{\text{tri}}$ are nearly 1/2 and 1/3 of their respective molecules indicates uniform intercalation of the molecules over their respective $s$ and implies that the TB pseudo-layers are spaced by the height of a single (tilted) alkyl-cyanobiphenyl or alkyl-biphenyl monomer unit making up the dimer or trimer. Interestingly, we find very little or no measurable temperature dependence of $s$ in the nematic and TB phases of the trimer. Likewise, Paterson *et al.* observe no discernable change in the equivalent $s_{\text{di}}$ feature in the dimer through the nematic and TB phases(14). For this reason, in the following analysis we consider $s$ to be constant through the temperature range of the TB phase.

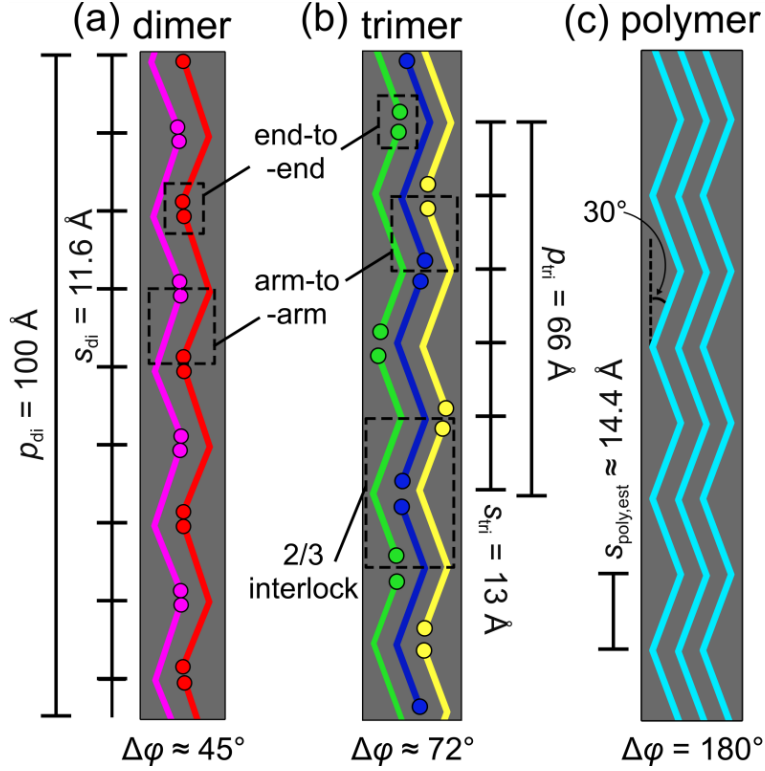

Fig. 3: Stick representations of the (a) CB6OCB dimer, (b) CB6OBO6CB trimer, and (c) the analogous theoretical polymer with corresponding low temperature TB parameters and azimuthal precession $\Delta\varphi$ set to 180° in each case for clarity. The molecules are uniformly intercalated, with the TB chains they form distinguishable and color-coded by the position of the chain-end interfaces. (a, b) The dimer and trimer systems both have end-to-end and arm-to-arm interactions, but only the trimer exhibits an additional 2/3 "interlocked" association (b) which tends to hinder molecular motion and flexibility in the TB helix. (c) The analogous polymer system forms an anticlinic (pseudo-)lamellar phase which is a special case of the TB phase with $\Delta\varphi = 180°$. The polymer molecules are constructed by attaching the ends of many rod-like units that make up the middle-monomer of the trimer (from the central carbon in one linker to the central carbon in the other). The pseudo-layer spacing $s_{poly,est}$ in the analogous polymer is estimated by tilting the middle rod-like unit of the trimer (16.6 Å) to a 30° angle.

From the values of the pitch $p$ and $s$, we can determine the azimuthal precession $\varphi$ per $s$ at low temperature to be $\Delta\varphi_{di} = 360° \times s_{di}/p_{di} = 360° \times (11.6\text{Å}/100\text{Å}) \approx 42°$ and $\Delta\varphi_{tri} = 360° \times s_{tri}/p_{tri} = 360° \times (13\text{Å}/66\text{Å}) \approx 71°$. For the sake of the geometric modelling that follows, we approximate by letting $\Delta\varphi_{di} \rightarrow 45°$ and $\Delta\varphi_{tri} \rightarrow 72°$. With this, we can construct the TB phase of the dimer as a series of octagons separated by the pseudo-layer spacing $s_{di}$, with the rod-like arms of the dimer lying along the diagonal of the rectangular faces of the octagonal prism formed by connecting the corners of the octagonal pseudo-layers with straight vertical lines, enforcing the azimuthal rotation of $\Delta\varphi_{di}$ per $s_{di}$ (Fig. 4a). Similarly, the TB phase of the trimer is modelled as a series of pentagons separated by $s_{tri}$, with the rod-like monomer units lying on the diagonal of the faces of the pentagonal prism formed, enforcing the azimuthal rotation of $\Delta\varphi_{tri}$ per $s_{tri}$ (Fig. 4b).

We can obtain the TB cone angle $\theta_{TB}$ from the optical birefringence measurements of the materials in the nematic and TB phases. On cooling from the nematic phase to the TB phase, we observe a decrease in birefringence, which is due to a collective tilt $\theta_{optical}$ of the molecules from $\hat{z}$ and can be determined by the method of Meyer *et al.*(23). This measured $\theta_{optical}$ is not, however, the TB cone angle $\theta_{TB}$, as $\theta_{optical}$ measures the tilt of the molecular *plane* of the molecule away from $\hat{z}$, and $\theta_{TB}$ is defined by the angle between the helix axis $\hat{z}$ and the local director $\hat{n}$. $\theta_{TB}$ comes from a combination of the intrinsic bend angle of the molecule *and* the tilting of the molecules from $\hat{z}$. We can therefore determine $\theta_{TB}$ at low temperature using $\theta_{optical}$ and the experimentally determined geometric constrains enforced by the polygonal constructions of Figs. 4a,b. By doing this, we find $\theta_{TB,di} = 21.2°$ ($\theta_{optical,di} \approx 20°$) and $\theta_{TB,tri} = 27.2°$ ($\theta_{optical,tri} \approx 17°$) at low temperature in the TB phase (see Supplementary Fig. S8 for derivation).

By solving our geometric constructions of the TB phase, we may also determine the effective monomer length $m_{eff}$ in the TB helix (the length of the diagonal of the rectangular face of the polygonal prisms in Figs. 4a,b). This length may in general be different than the full length of the rod-like monomer unit $m_{calc}$ making up the dimer and trimer, if there is overlap or separation between end groups, for instance. Solving from the geometric constraints in the TB model of the dimer (Fig. 4a), we find $m_{di,eff} = 12.4$ Å. The length of one of the rod-like monomers making up the dimer, as measured in its extended all-*trans* conformation from the central carbon in the linker to the end nitrogen group, is calculated to be $m_{di,calc} = 15.8$ Å. Here, we consider only the extended all-*trans* conformer of the dimer as an approximation to the conformational diversity expected in the TB phase, since the condensed and locally orientationally aligned TB phase tends to favor extended conformations over more kinked conformers. This indicates an overlapping of the end groups for the dimer of $m_{di,calc} - m_{di,eff} = 3.4$ Å at low temperature in the TB phase, or a bit more than the atomic diameter of an end nitrogen. We expect that on heating, in addition to the cone angle decreasing, the ends of the dimers will tend to pull apart, leading to the flexibility in the helix structure and the dramatic pitch increase that we observe at high temperatures in the TB phase (Fig. 2a).

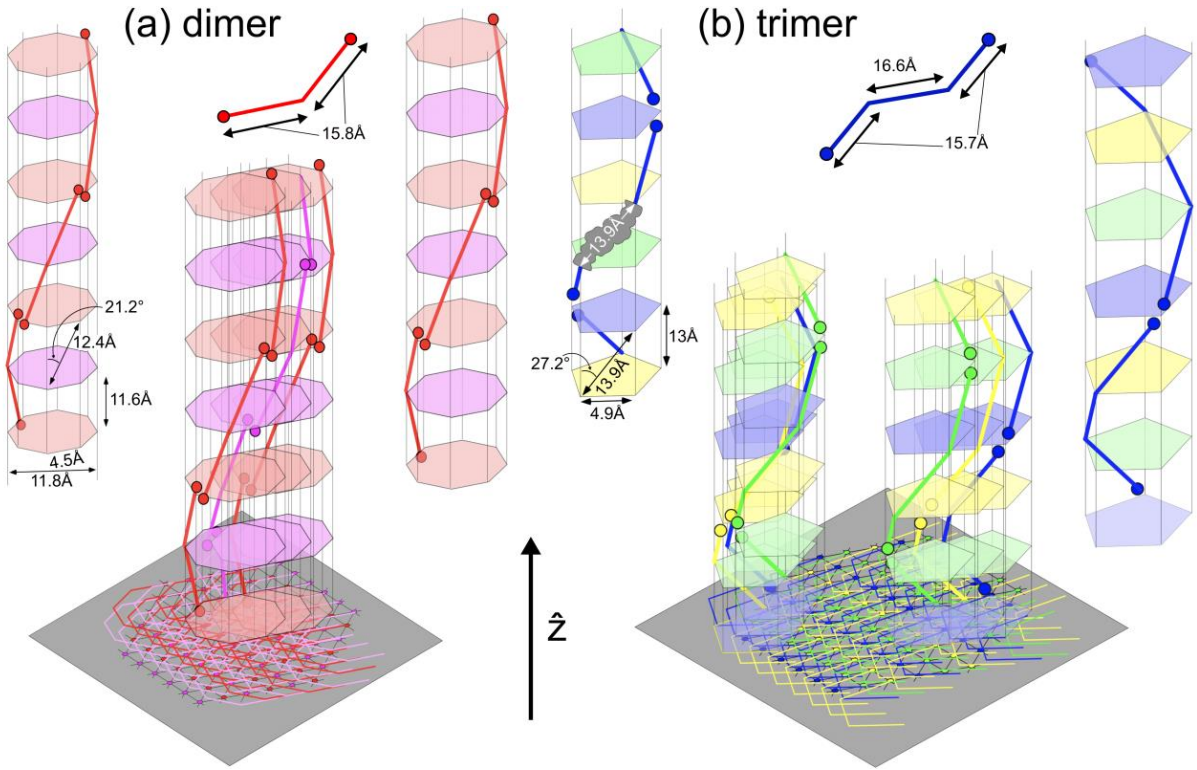

Fig. 4: Geometric models of the CB6OCB dimer and CB6OBO6CB trimer in their respective TB phases. (a) A geometric construction that governs the structure of the TB phase formed by the dimer at low temperature. The chains are divided into 2 subsets, with the ends of the red molecules in the chains meeting at the corner of a red octagon, and the ends of the purple molecules meeting at the corner of a purple octagon. The molecular ends in the dimer overlap by 15.8 – 12.4 Å = 3.4 Å, and the rod-like arms are tilted from the helix axis ẑ along the diagonal of an outer rectangular face of the octagonal prism by the TB cone angle $\theta_{TB,di} = 21.2°$. (b) A geometric construction governing the structure of the TB phase formed by the trimer at low temperature. The chains are divided into 3 subsets with the ends of the blue molecules in the chains meeting at a blue pentagon, and likewise for the green and yellow chains. The chemical make-up of a rod-like middle-monomer is depicted in grey overlaying a middle segment of the blue leftmost trimer chain. The molecular ends of the trimer overlap by 16.0 – 13.9 Å = 2.1 Å (16.0 Å being the averaged monomer length), which is significantly less than that found for the dimer. The rod-like units of the trimer are tilted along the diagonal of the outer face of the pentagonal prism by the cone angle $\theta_{TB,tri} = 27.2°$. The polygonal prisms in (a) and (b) enforce the respective azimuthal precessions $\Delta\varphi$ per pseudo-layer *s*. Once the single chains are constructed, they may be rotated around the center line of the polygons to change their helical phase. In this 3D packing sketch, the phase is assumed to be the same for all helices, with the differently colored helices positioned randomly in their packing. We depict a possible 3D packing of the helical chains on a hexagonal lattice in (a) and (b) (grey planes), though the actual packing motif may be different.

Using a similar treatment for the trimer as for the dimer, we determine the effective monomer length in the TB phase of the trimer to be $m_{tri,eff} = 13.9$ Å. Because the end- and middle-monomer groups of the trimer are not of equal lengths, we can determine the average monomer length using the Spartan calculation of the extended all-*trans* conformation of the trimer by averaging the lengths of the individual monomer segments. We do this because we only observe a single scattering peak from the pseudo-layer spacing, which implies that the end- and middle-monomer units that make up the trimer are uniformly distributed in 3D space. Here again, we consider the extended all-*trans* conformer of the trimer in our model as an approximation, for the same reasons that we do for the dimer (as mentioned above). The length of an end-monomer is measured from the central carbon in a carbon linker to its nearest nitrogen atom, and the length of a middle-monomer is measured from the central carbon of one alkyl linker to the central carbon of the other linker. From this, we find $m_{tri,calc} = (2\times15.7\text{Å}+16.6\text{Å})/3 = 16\text{Å}$. And now, $m_{tri,calc} - m_{tri,eff} = 2.1$ Å, indicating a smaller overlap of the molecular ends in the trimer than that found in the dimer. Indeed, the mismatch in the lengths of the end- and middle-monomers of the trimer will tend to pull the shorter end-monomers apart by association with the longer middle-monomers (Fig. 3b). From the above discussion, we can conclude that the molecules in the TB phase of the dimer and the trimer exhibit mutual end group interactions which leads to an effective heliconical chain-like arrangement of the molecules, as depicted in Figs. 4a,b.

Now we turn to the possible intermolecular interactions which occur in the TB phases of the dimer and trimer. In the dimer, there are two kinds of possible interactions, given that the molecules are uniformly intercalated over half their length: (1) the ends of different dimers associate, and (2) the molecular arms of different dimers associate (Fig. 3a). Both the end-to-end and the arm-to-arm associations are weak physical connections, as they are only held together by some combination of van der Waals and electrostatic forces, thereby contributing to flexibility, elasticity, and fluctuations in the structure of the TB helix. This is especially evident at higher temperature in the TB phase, such that the pitch varies significantly with temperature in the dimer as we observe.

The trimer system exhibits an additional molecular association which contributes to a more rigid TB helix structure, as evidenced in Fig. 2b. The uniformly intercalated trimers form the TB helix with 3 distinct molecular associations: (1) the ends of different trimers associate, (2) the end molecular arms of different trimers associate, and (3) 2/3 of neighboring trimer molecules will overlap, or "interlock" together (Fig. 3b). These 2/3 interlocks sterically hinder the motion of the interlocked molecules and their neighbors. This additional interaction is a result of more chemically-bonded monomers per unit length in the TB helix of the trimer than there are in the dimer. For the trimer, at each pseudo-layer interface (any pentagon in Fig. 4b) there are two-thirds chemically bonded monomers (chemically bonded with an alkyl linker) and one-third terminally associated monomers (end-to-end physical association), whereas for the dimer, at each pseudo-layer interface (any octagon in Fig. 4a) there are one-half chemically bonded monomers and one-half terminally associated monomer units. This effect tends to increase the rigidity of the TB helix thereby reducing (and essentially eliminating) the temperature-dependence of the helix pitch in the TB phase of the trimer.

From our experiments, a number of interesting trends in the structure of the TB phase of this oligomeric family emerge. For instance, the pseudo-layer spacing increases from the dimer to the trimer ($s_{di} = 11.6$ Å; $s_{tri} = 13$ Å). This trend indicates that $s$ will increase in the higher oligomers, but is not likely to exceed the theoretical pseudo-layer spacing $s_{poly,est} \approx 14.4$ Å of the comparable polymer which is composed of the middle-monomer units of the trimer tilted at a comparable angle (30°), as depicted in Fig. 3c. This is because the comparable polymer, in the anticlinic (pseudo-) lamellar system, has (effectively) no molecular ends that may overlap, in contrast to the dimer and trimer which have a mixture of inner and terminal connections between rod-shaped elements which do overlap. In addition, the TB pitch tends to decrease from the dimer to the trimer ($p_{di}$(low $T$) $\approx$ 100 Å; $p_{tri}$(low $T$) $\approx$ 66 Å), implying that the pitch will tend to decrease in the higher oligomers in this family and approach a value not smaller than $p_{poly} = 2 \cdot s_{poly,est}$. The pitch will also become more temperature-independent with the increasing number of monomers in the oligomer as the amount of intermolecular locks increases. Additionally, the azimuthal precession $\Delta\varphi$ per $s$ increases from the dimer to the trimer ($\Delta\varphi_{di} \approx 42°$; $\Delta\varphi_{tri} \approx 71°$ at low $T$). From this observation, we can expect that $\Delta\varphi$ will tend to increase in the higher oligomers of this family, and asymptotically approach a value perhaps not exceeding $\Delta\varphi_{poly} = 180°$, since the comparable polymer in an anticlinic (pseudo-)lamellar phase, has $p_{poly}$ and $s_{poly}$, (and therefore $\Delta\varphi_{poly}$), fixed by the chemical structure. Our investigation of this dimer and trimer outlines the possible trends in their oligomeric family and for other linear main-chain oligomeric series with similar molecular constructions. It is important to stress, however, that the behavior of the TB nanostructure of different oligomer families may depend qualitatively on the chemical structure of those oligomers, as well as on the number of monomer segments, their relative dimensions, and preferred orientations. Clearly, it is of great interest to explore the features of the TB nanostructure more systematically as a function of these variables.

Finally, we evaluate the possible 3D packing motifs of the heliconical chains formed from the dimer and trimer. Consider the hexagonal lattice represented in the grey shaded planes in Figs. 4a,b. Each site on this lattice has a heliconical chain passing through it, and the constructions in Figs. 4a,b result if we assume that all the chains pass through the plane with the same phase $\varphi$. The assumption that all the chains have the same phase is motivated by the x-ray diffraction crystal structures of Hori *et al.*(27) for their dimer material II-3, which exhibits a helical state in its crystal phase with an azimuthal rotation of $\Delta\varphi_{\text{II-3}} = 90°$ per layer, with a layer spacing of $s_{\text{II-3}} = 9.7$ Å, a tilt of the mesogenic cores of $\theta_{\text{II-3}} = 45°$ relative to the helix axis, and a pitch of $p_{\text{II-3}} = 38.7$ Å. The crystal structure of II-3 is similar to that sketched for the TB phase of the trimer in Fig. 4b, but with II-3 having its helical chains on squares rather than on pentagons and tiling a square lattice to fill 3D space. In this crystal structure, the helical phase $\varphi$ in a given plane normal to the helix axis is the same for all the chains, a geometrical consequence of the large monomer tilt of II-3 and the steric constraints of packing the rod-shaped units with a large tilt. The large inherent tilt of the rod-like monomer units in the dimer and trimer makes it likely that their helical chains are likewise precessing in phase with one another, as depicted in Figs. 4a,b. Though we find that this evidence points to the organization depicted in Figs. 4a,b, it is possible that the helical chains are instead randomly spatially arranged or arranged in more complex fashions, though we cannot speak to these organizational motifs in the present study.

Recently, Al-Janabi *et al*. published an investigation of the phase behavior of the trimer that we study here(8). A comparison of our data for our trimer and theirs (which they call B6$_3$) show that their differential scanning calorimetry measurements, phase diagram, and non-resonant x-ray scattering features are nearly the same as ours. The only difference appears to be that they obtained different optical textures than expected for a TB phase. On this basis, and the lack of miscibility between their trimer and another material with a TB phase, they chose not to identify the lower temperature nematic phase as a TB phase. However, as we report, our RSoXS experiments demonstrate an orientation modulation at 66 Å only near the carbon K-edge resonance, and not off the energy resonance (see Supplementary Fig. S4). The TB phase is the only known LC phase in this range of length-scales (and, correspondingly, $q$-values) that exhibits resonant scattering for $p$ and the absence of scattering for $p$ in non-resonant conditions. The likelihood of having a splay-bend nematic or similar modulated nematic with a non-zero density modulation, but lacking a non-resonant scattering signal is too remote to consider and has already been discussed in detail in ref.(13). In addition to this evidence, our polarized light microscopy images are decidedly typical of the TB phase, exhibiting the blocky texture and the rope-texture in a unidirectionally rubbed cell, with the textures alone having been used by many research groups to identify the TB phase. And finally, a binary phase diagram (Supplementary Fig. 2) and a contact cell preparation with the trimer and the well-known TB-forming material CB7CB (Supplementary Fig. 3) clearly demonstrate miscibility of these compounds in the nematic and TB phases. It is with this evidence that we confidently identify the lower-temperature nematic phase of CB6OBO6CB as the TB phase.

## CONCLUSIONS

We synthesized and characterized the TB-forming LC trimer CB6OBO6CB and its analogous dimer CB6OCB. The trimer exhibits a monotropic TB phase with a temperature-independent heliconical pitch $p_{tri} = 67 – 66$ Å, the smallest yet reported in the literature. Our characterization and analysis of these materials suggest the importance of the number of chemically linked segments in a molecule for controlling the elasticity and fluctuations in the TB nanostructure. We constructed a geometric picture of the TB phase for the dimer and trimer which indicates that they are formed by a helical chain motif which is linked together by an association and overlapping of the molecular ends. The nanoscale properties of their TB phases differ significantly because the trimer exhibits an additional molecular association over 2/3 its length which does not exist in the dimer. This work evokes the possibility for designing LCs which exhibit a specific temperature-independent TB pitch for technological applications in which a fixed pitch over a wide temperature range is preferred, such as nano-templating or chiral separation techniques. It also opens up a rich phase-space of possible molecular designs motifs for the TB phase by not only varying the chemical make-ups and number of monomer units, but by also varying the relative physical dimensions and orientations of the monomers in the oligomer, likely yielding further untold exciting material properties.

## MATERIALS AND METHODS

The synthesis of CB6OBO6CB is described in detail in the ESI.

The thermal behavior of CB6OBO6CB was investigated by differential scanning calorimetry using a Mettler Toledo DSC822$^e$ differential scanning calorimeter equipped with a TSO 801RO sample robot and calibrated using indium and zinc standards. The heating profile in each run was heat, cool, and reheat at 10°C min$^{-1}$ with a 3-min isotherm between heating and cooling segments. Thermal data were normally extracted from the second heating trace.

Initial phase characterization was performed with polarized light microscopy, using an Olympus BH2 polarizing light microscope equipped with a Linkam TMS 92 hot stage.

Birefringence measurements of CB6OCB and CB6OBO6CB were carried out using polarized light microscopy, with the samples filled into 4.2 μm thick unidirectionally-rubbed glass cells. The birefringence at a given temperature was determined using the average of 3 distinct measurements from a Berek optical compensator.

RSoXS experiments were performed at the Advanced Light Source at Lawrence Berkeley National Laboratories beamline 11.0.1.2 with linearly polarized x-ray photons. RSoXS cells were assembled by placing a 100 nm silicon nitride window, supported by a silicon substrate on a hot plate with the temperature above the clearing temperature. The LC material was placed onto the substrate, with another identical substrate placed on top of the material to form a sandwich cell. This cell was mounted onto a custom-made copper hot stage, which permitted heating the sample in situ. The hot stage was placed inside a vacuum chamber (P <10$^{-6}$ Torr) due to the extremely short attenuation length of soft X-rays in air. The x-ray energy was tuned between 270 eV and 290 eV in our experiments. We took detector images of CB6OCB and CB6OBO6CB on cooling from the nematic phase. The data from the detector was reduced using the Igor Pro-based NIKA data reduction software package(28, 29).

WAXS experiments were performed at the Advanced Light Source at Lawrence Berkeley National Laboratories beamline 7.3.3(30). We filled the LC into thin borosilicate capillaries of 1 mm diameter, which were mounted inside an Instec hot stage. The hot stage was sealed with Kapton tape, which contributes a background signal at $q \approx 0.4$ Å$^{-1}$. The incident beam energy was 10 keV.

Freeze-fracture transmission electron microscopy (FFTEM) experiments were carried out by sandwiching the LC between 2 mm × 3 mm glass planchettes, which induce mostly random planar anchoring of the molecules at the glass interfaces. We cooled the cell from the isotropic (I) to the desired LC phase and monitored the sample in the microscope. The cell was then rapidly quenched to $T < -180°C$ by immersion in liquid propane and fractured under high vacuum at $-140°C$. The exposed LC surface was subsequently coated with 2 nm of platinum deposited at 45° for imaging contrast, followed by ~25 nm of carbon deposited at 90° to increase the mechanical rigidity of the replica. After removing the LC material, the Pt–C replica was placed on a copper TEM grid and imaged in a Philips CM 10 100 keV TEM, allowing the topography of the fracture plane to be observed. TEM images were obtained with a 1K × 1K Gatan Bioscan digital camera. The surfaces facing the platinum shadowing direction accumulate more platinum and therefore produce darker shadows in the TEM images.

We estimated the end-to-end length of CB6OBO6CB trimer and CB6OCB dimer in their extended *all-trans* conformations. Dihedrals connecting the rings to the alkyl chains were initially manually set at their expected energy minima—values of 0 degrees for CR-CR-O-CT and 90 degrees for CR-CR-CT-CT (where CR is an aromatic ring carbon atom, O is an oxygen atom and CT the tetrahedral alkyl chain carbon). The trimer and the dimer were optimized to reach an equilibrium geometry, in vacuum in the ground state, using the semi-empirical AM1 method as an initial approximation. Subsequently, a second fully relaxed geometry optimization we initiated using the *ab initio* Hartree-Fock method with 6-31G* basis set.

Molecular end-to-end lengths for both the trimer and dimer were then determined by measuring the distance from one terminal nitrogen atom to the other and adding the length of 2 nitrogen radii. The end-to-end nitrogen-nitrogen distance of the trimer was $l_{tri} = 46.0$ Å, and was $l_{di} = 30.4$ Å for the dimer, very close to that estimated for the dimer in ref. (14). We then determined the end-to-end molecular dihedral angles of the trimer by comparing one nitrogen-oxygen plane to the other nitrogen-oxygen plane, finding it to be 134.08 degrees. To find the dihedral angle of the dimer, we compared the nitrogen-oxygen plane to a $C_b$-nitrogen plane (where $C_b$ is the benzylic carbon attached to the cyanobiphenyl unit) and determined it to be 0.28 degrees.

All molecular modeling calculations were performed with Spartan'16 molecular calculation and modeling software.

Corresponding author emails: michael.tuchband@colorado.edu, chenhuizhu@lbl.gov, c.t.imrie@abdn.ac.uk, ajakli@kent.edu, noel.clark@colorado.edu

The authors declare no competing financial interest

**ACKNOWLEDGEMENTS**

This work was supported by National Science Foundation Materials Research Science and Engineering Center Grant DMR-1420736 and NSF Grant DMR-1307674. MRT acknowledges support from the Advanced Light Source doctoral fellowship in residence offered by Lawrence Berkeley National Laboratories. MS acknowledges the support of the U.S. National Science Foundation I2CAM International Materials Institute Award, Grant DMR-1411344. We acknowledge use of beamlines 11.0.1.2 and 7.3.3. of the Advanced Light Source supported by the Director of the Office of Science, Office of Basic Energy Sciences, of the U.S. Department of Energy under contract no. DE-AC02-05CH11231.

SUPPLEMENTARY INFORMATION

### SECTION 1: MATERIALS/ GENERAL METHODS/ INSTRUMENTATION

For reactions performed under anhydrous conditions, all glassware was pre-dried for at least 12 h in ovens set at 120 °C.

**Materials**

All reagents and solvents were available commercially and purchased from Sigma Aldrich, TCI Chemicals or Alfa Aesar and were used as received unless otherwise stated. Anhydrous solvents were purchased as anhydrous (over molecular sieves).

Solvents were evaporated at approximately 20 mm Hg using a water aspirator pump connected to a Buchi rotary evaporator and trace solvents in a Thermo Scientific vacuum oven at 1.0 mm Hg and 50 °C.

Column chromatography was performed using silica gel grade 60A 40-63 µm, purchased from Flurochem and a small neutral alumina plug was used at the base of the column to remove ionic impurities where stated. Reactions were monitored using Thin Layer Chromatography (TLC) carried out on aluminium-backed plates with a coating of Merck Kieselgel 60 F254 silica and an appropriate solvent system. Silica gel coated aluminium plates were purchased from Merck KGaA. Spots were visualised using UV light (254 nm) or by oxidation with either an aqueous permanganate dip or iodine.

**General methods and instrumentation**

Infrared spectra were recorded on a Thermo Scientific Nicolet IR100 FT-IR spectrometer with an ATR diamond cell.

Mass spectra were recorded on a Waters QTOF Xevo G2 spectrometer.

Proton ($^1$H) and carbon ($^{13}$C) NMR spectra were recorded on a Varian Unity INOVA 600 MHz NMR spectrometer, a Varian Unity INOVA 400 MHz NMR spectrometer or a 300 MHz Bruker Ultrashield NMR spectrometer. The chemical shifts $\delta$ are quoted in parts per million (ppm) (SiMe$_4$, $\delta = 0$), using residual non-deuterated solvent signals as reference. Coupling constants ($J$ values) are quoted in Hertz (Hz) and are vicinal $^3J$, unless otherwise indicated. The splitting patterns are reported using the following abbreviations: b (broad), s (singlet), d (doublet), t (triplet), q (quartet), quin (quintet), m (multiplet), and combinations thereof. $^{13}$C Spectra are proton decoupled unless otherwise stated. Ar refers to an aromatic ring.

The purity of final products were verified using C,H,N microanalysis performed by the Micro Analytical Laboratory in the School of Chemistry at the University of Manchester or the Centre for Chemical Instrumental Analysis and Services at the University of Sheffield.

## SECTION 2: SYNTHETIC PROCEDURES

## Synthesis of 4,4'-bis((6-(4'-cyano-[1,1'-biphenyl]-4-yl)hexyl)oxy)-1,1'-biphenyl, CB6OBO6CB

The synthetic route used to obtain 4,4'-bis((6-(4'-cyano-[1,1'-biphenyl]-4-yl)hexyl)oxy)-1,1'-biphenyl, CB6OBO6CB, is shown in scheme 1.

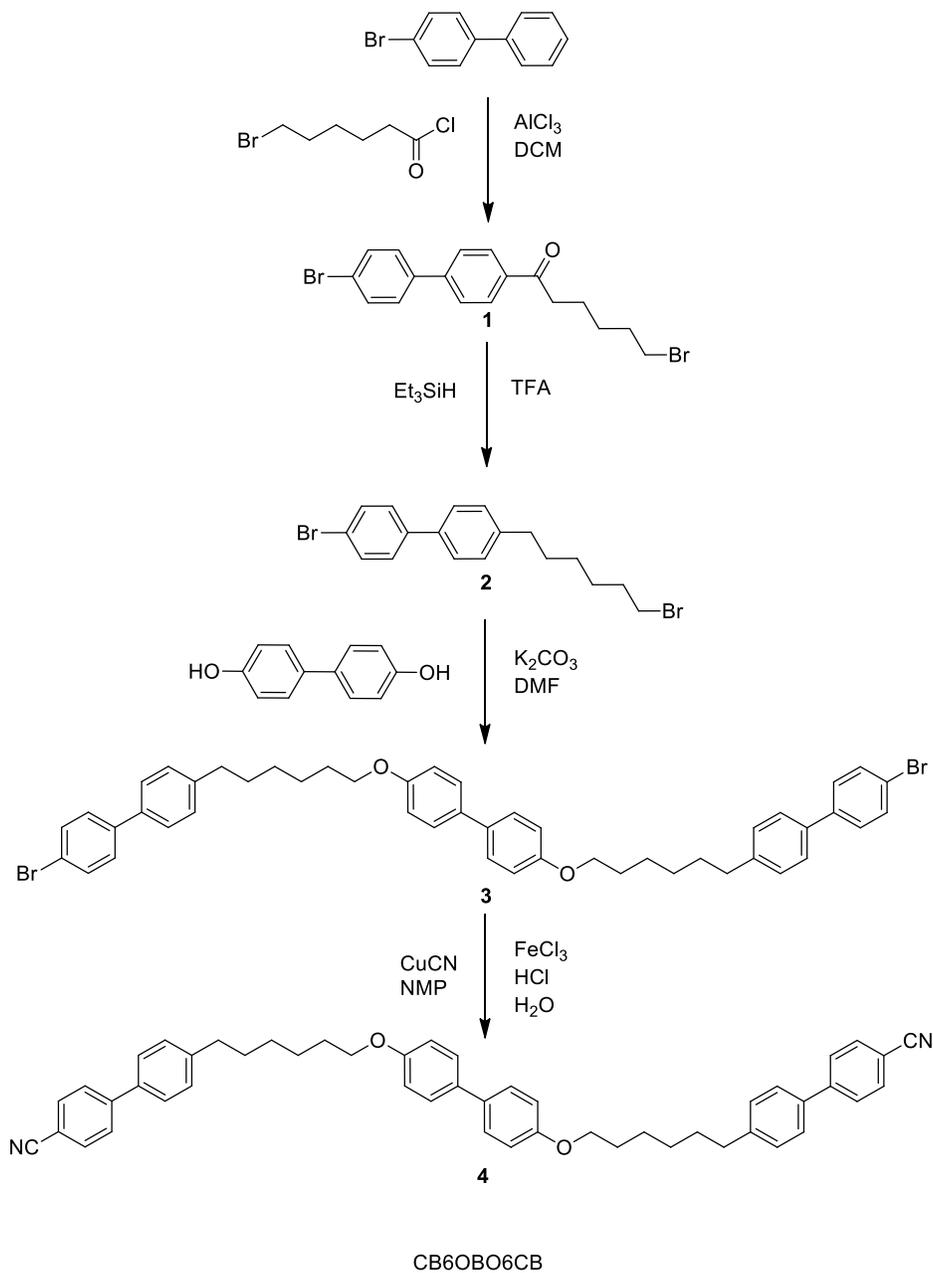

**Scheme 1.**

### 2.1.1. 4-Bromo-4'-(6-bromohexanoyl)biphenyl, *1*

The synthesis of compound **1** has been described in detail elsewhere(14).

### 2.1.2. 1-Bromo-6-(4'-bromobiphenyl-4-yl)hexane, *2*

Compound **2** was prepared by reduction of the carbonyl group on **1** using triethylsilane in trifluoroacetic acid(31) and has been fully described elsewhere(14).

### 2.1.3. 4,4'-bis((6-(4'-bromo-[1,1'-biphenyl]-4-yl)hexyl)oxy)-1,1'-biphenyl, *3*

A mixture of **2** (4.67 g, 0.012 mol), 4,4'dihydroxybiphenyl (1.02 g, 0.0056 mol), potassium carbonate (3.17 g, 0.022 mol) and dimethylformamide (50 mL) was heated at reflux overnight. The reaction mixture was cooled to room temperature, poured into $H_2O$ (150 mL), and the white precipitate was collected by vacuum filtration. The crude product thus obtained was recrystallized from ethyl acetate to give the title compound as a white solid. Yield: 3.04 g, 33%. $T_{CrI}$ 173 °C. Infrared ν cm$^{-1}$: 2921 (C-H), 1606, 1499, 1480, 1275, 1247, 999, 826, 807, 518, 473. $^1$H NMR (300 MHz CDCl$_3$) δ: 7.56 (4H, d, *J* 8.6 Hz, Ar), 7.47 (12H, m, Ar), 7.27 (4H, d, *J* 8.6 Hz, Ar), 6.95 (4H, d, *J* 8.6 Hz, Ar), 4.00 (4H, t, *J* 6.4 Hz, OCH$_2$CH$_2$), 2.68 (4H, t, *J* 7.7 Hz, ArCH$_2$CH$_2$), 1.83 (4H, quin, *J* 7.1 Hz, OCH$_2$CH$_2$CH$_2$), 1.71 (4H, quin, *J* 7.6 Hz, ArCH$_2$CH$_2$CH$_2$), 1.53 (4H, m, CH$_2$CH$_2$CH$_2$CH$_2$), 1.48 (4H, m, CH$_2$CH$_2$CH$_2$CH$_2$). $^{13}$C NMR (151 MHz, CDCl$_3$) δ 158.33, 142.47, 140.19, 137.50, 133.48, 131.94, 129.12, 128.70, 127.79, 126.94, 121.31, 114.90, 77.37, 77.16, 76.95, 68.10, 35.61, 31.46, 29.35, 29.08, 26.06.

### 2.1.4. 4,4'-bis((6-(4'-cyano-[1,1'-biphenyl]-4-yl)hexyl)oxy)-1,1'-biphenyl, *4*, CB6OBO6CB

The cyanation of **3** was achieved using a modified Rosenmund-von Braun reaction as described by Coates and Gray(32). A mixture of **3** (3.04 g, 0.004 mol), copper cyanide (1.30 g, 0.015 mol) and dry N-methyl-2-pyrrolidone (100 mL) was heated to 200 °C for 24 h. The reaction mixture was cooled to 80 °C and to this was added a solution of iron chloride (6.90 g, 0.05 mol), $H_2O$ (15 mL) and 32 % aq. hydrochloric acid (6 mL) at 60 °C. This solution was allowed to cool slowly to room temperature and stirred overnight, then added to a dichloromethane (200 mL) and H$_2$O (200 mL) mix. The aqueous layer was washed with dichloromethane (100 mL). All organic fractions were combined and washed with H$_2$O (3 x 100 mL) before drying over anhydrous magnesium sulfate. Solvent was removed under vacuum to yield a brown liquid which was added to H$_2$O (200 mL). The brown precipitate was collected by vacuum filtration and washed with H$_2$O (400 mL). The crude product was purified by silica gel chromatography using dichloromethane as eluent, R.f. 0.54. The crude product thus obtained was recrystallized from ethanol to give the title compound as a white solid. Yield: 1.23 g, 47%. T$_{CrN}$ 140 °C, T$_{N_{TB}N}$ 131 °C, T$_{NI}$ 169 °C. Elemental analysis: Calculated for C$_{50}$H$_{48}$N$_2$O$_2$: C, 84.71%, H, 6.82%, N, 3.95%. Found: C, 84.36%, H, 7.02%, N, 4.00%. Infrared ν cm$^{-1}$: 2924 (C-H), 2853 (C-H), 2226 (C≡N), 1604, 1496, 1473, 1244, 1174, 816, 797, 597, 517. $^1$H NMR (300 MHz CDCl$_3$) δ: 7.73 (4H, d, *J* 8.6 Hz, Ar), 7.68 (4H, d, *J* 8.6 Hz, Ar), 7.52 (4H, d, *J* 8.6 Hz, Ar), 7.47 (4H, d, *J* 8.6 Hz, Ar), 7.32 (4H, d, *J* 8.6 Hz, Ar), 6.96 (4H, d, *J* 8.6 Hz, Ar), 4.01 (4H, t, *J* 6.1 Hz, OCH$_2$CH$_2$), 2.71 (4H, t, *J* 7.6 Hz, ArCH$_2$CH$_2$), 1.84 (4H, quin, *J* 7.2 Hz, OCH$_2$CH$_2$CH$_2$), 1.72 (4H, quin, *J* 7.5 Hz, ArCH$_2$CH$_2$CH$_2$), 1.56 (4H, m, CH$_2$CH$_2$CH$_2$CH$_2$), 1.47 (4H, m, CH$_2$CH$_2$CH$_2$CH$_2$). $^{13}$C NMR (75 MHz CDCl$_3$) δ: 158.19, 145.59, 143.56, 136.52, 133.31, 132.58, 129.22, 127.66, 127.49, 127.11, 119.08, 114.73, 110.53, 67.90, 35.49, 31.27, 29.19, 28.90, 25.90. MS (ESI+, m/z): [M+Na]$^+$ Calculated for C$_{50}$H$_{48}$N$_2$O$_2$Na: 731.3613. Found: 731.3605.

**SECTION 3: SUPPLEMENTARY FIGURES**

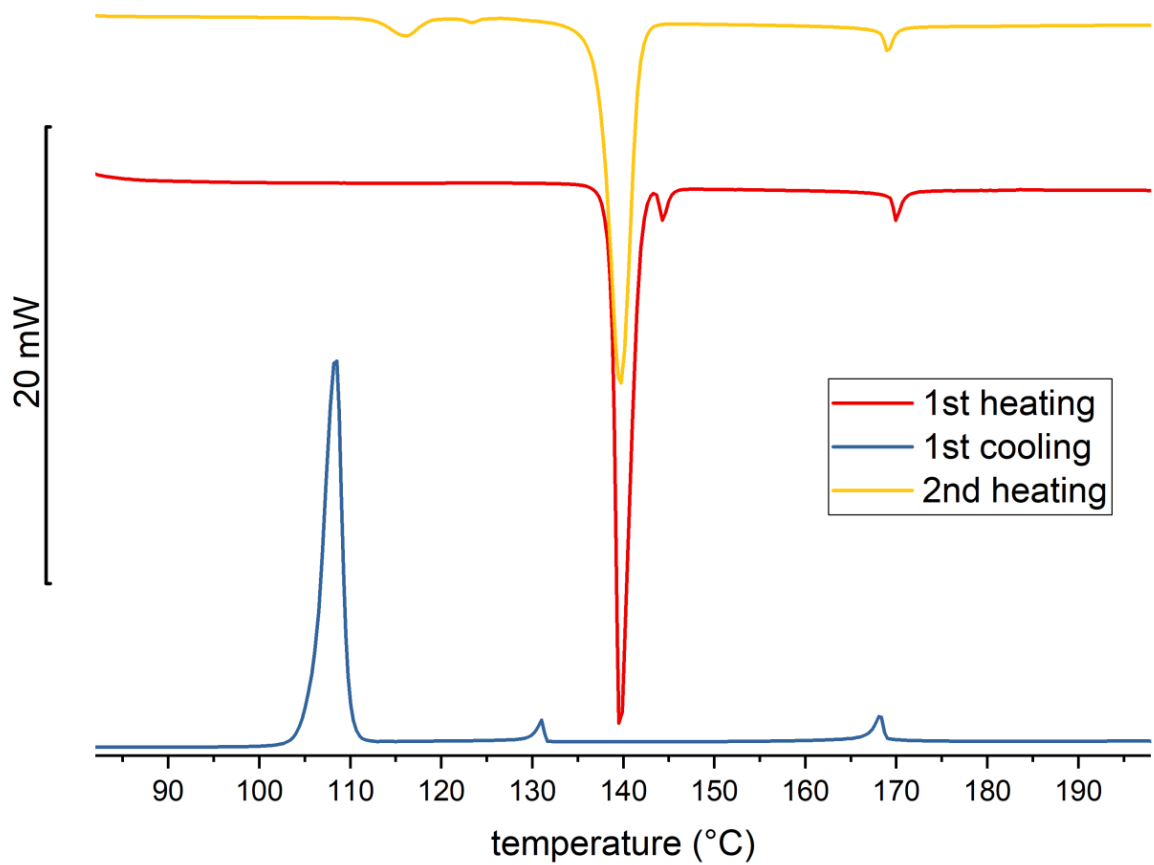

| molecule | $T_{Cr}$ (°C) | $T_{TB-N}$ (°C) | $T_{N-I}$ (°C) | $\Delta H_{Cr}$ (kJ mol$^{-1}$) | $\Delta H_{TB-N}$ (kJ mol$^{-1}$) | $\Delta H_{N-I}$ (kJ mol$^{-1}$) | $\Delta S_{Cr}/R$ | $\Delta S_{TB-N}/R$ | $\Delta S_{N-I}/R$ |
|---|---|---|---|---|---|---|---|---|---|
| CB6OCB | 99 | 109 | 155 | 21.29 | 0.02 | 1.58 | 6.88 | 0.01 | 0.48 |
| CB6OBO6CB | 140 | 131* | 169 | 64.51 | 1.33 | 1.81 | 18.78 | 0.40 | 0.49 |

Supplementary Fig. S1: Differential scanning calorimetry plots of CB6OBO6CB on heating, cooling, and reheating, and associated thermodynamic parameters.

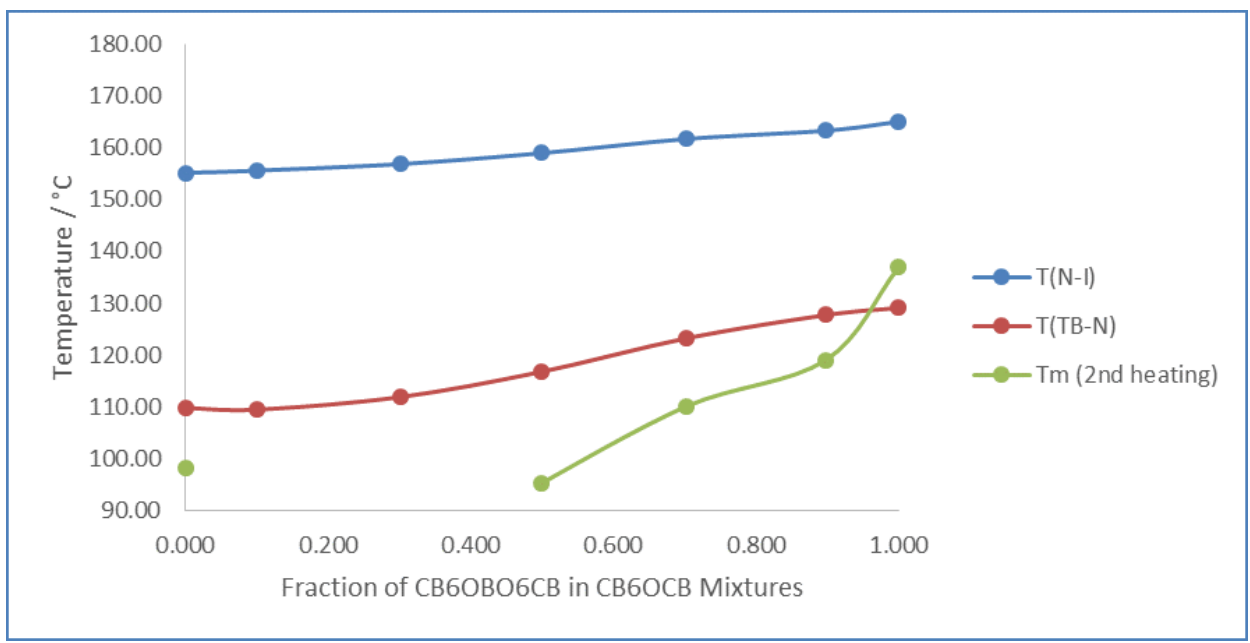

Supplementary Fig. S2: Binary phase diagram of CB6OBO6CB (trimer) and CB6OCB (dimer) in concentration and temperature, demonstrating good mixing between the nematic and TB phases of each component. This shows that the phase below the nematic is indeed the same phase for both the dimer(14) and the trimer: the TB phase.

Supplementary Fig. S3: Contact preparation of CB7CB and CB6OBO6CB. The polarizers are oriented along the edges of the images. The images taken at $T = 138.5 – 118.5°C$ have 1.0 mm vertical dimension, while the images taken at $T = 117.0 – 67.6°C$ have 1.7 mm vertical dimension.

The CB7CB-rich region is toward the upper-left portion of the images, and the trimer-rich regions are on the bottom-right. At $T = 138.5°C$, CB7CB is isotropic, while the trimer is in the nematic phase. The region with the parallel line defects is the region of significant mixing and is nematic at this temperature. On cooling ($T = 132.1°C$), the I – N phase front continues to grow as the trimer-rich region transitions to the TB phase and exhibits the characteristic blocky texture. For $T = 130.4 – 118.5°C$, both the I – N and N – TB phase fronts progress through the mixed region (from the $T = 118.5°C$ image, the field of view is shifted to the left by ~500 μm). At $T = 117.0°C$, the CB7CB-rich region transitions to the nematic phase. From $T = 115.3 – 105.6°C$, the field of view is again shifted ~500 μm to the left and the N – TB phase front continues to grow steadily. At $T = 104.7°C$ and $104.0°C$, the CB7CB-rich region transitions to the TB phase, exhibiting the usual blocky textures. The continuity of the I – N and the N – TB phase front from the trimer-rich region to the CB7CB-rich region demonstrates that the materials both exhibit the nematic and TB phases and are well-mixed in these phases. At $T = 102.2°C$ and $T = 67.6°C$, we shift the field of view to the right by ~500 μm to observe the lower temperature TB-Cr phase front grown in from the trimer-rich region. This experiment demonstrates that the monotropic LC phase exhibited by the trimer from $T = 130 – 110°C$ is indeed the usual TB phase.

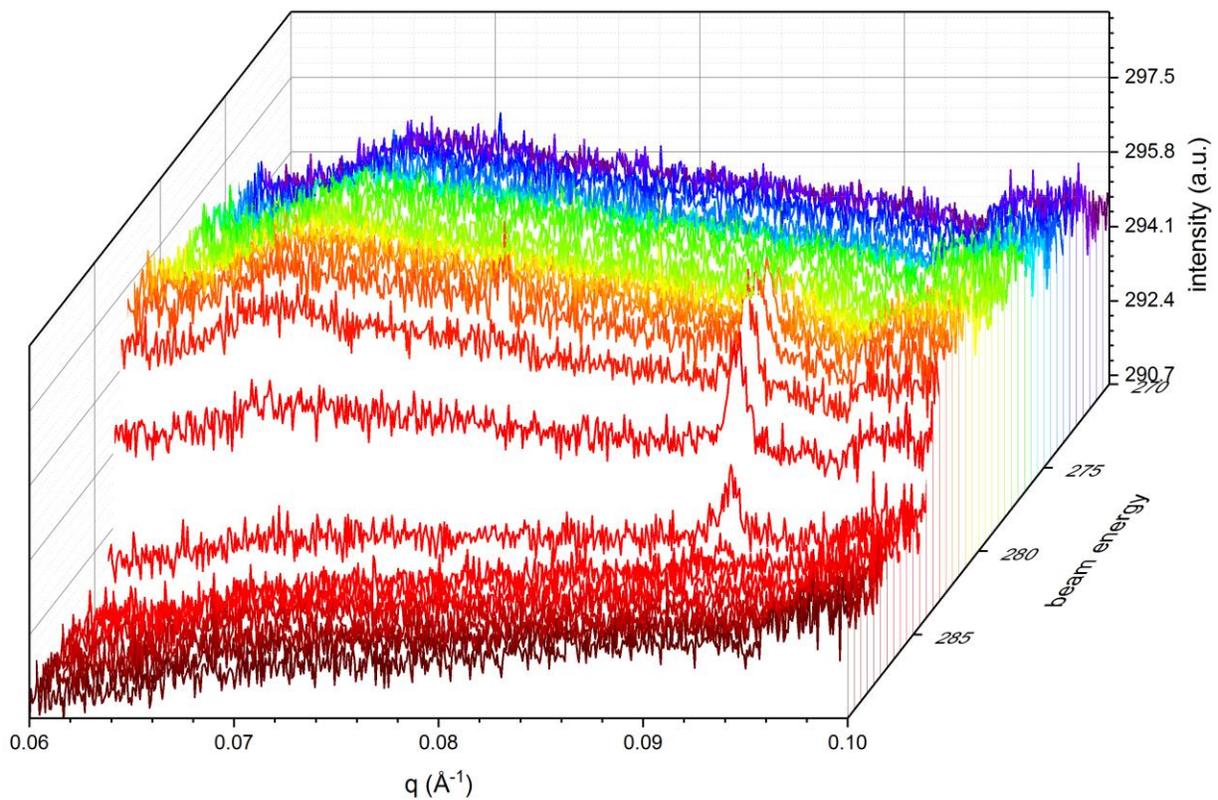

Supplementary Fig. S4. Beamline energy scan of CB6OBO6CB. The peak corresponding to the TB phase at $q_{tri} \approx 0.095$ Å$^{-1}$ becomes observable in the vicinity of the carbon K-edge resonance ($E_{res} = 284$ eV).

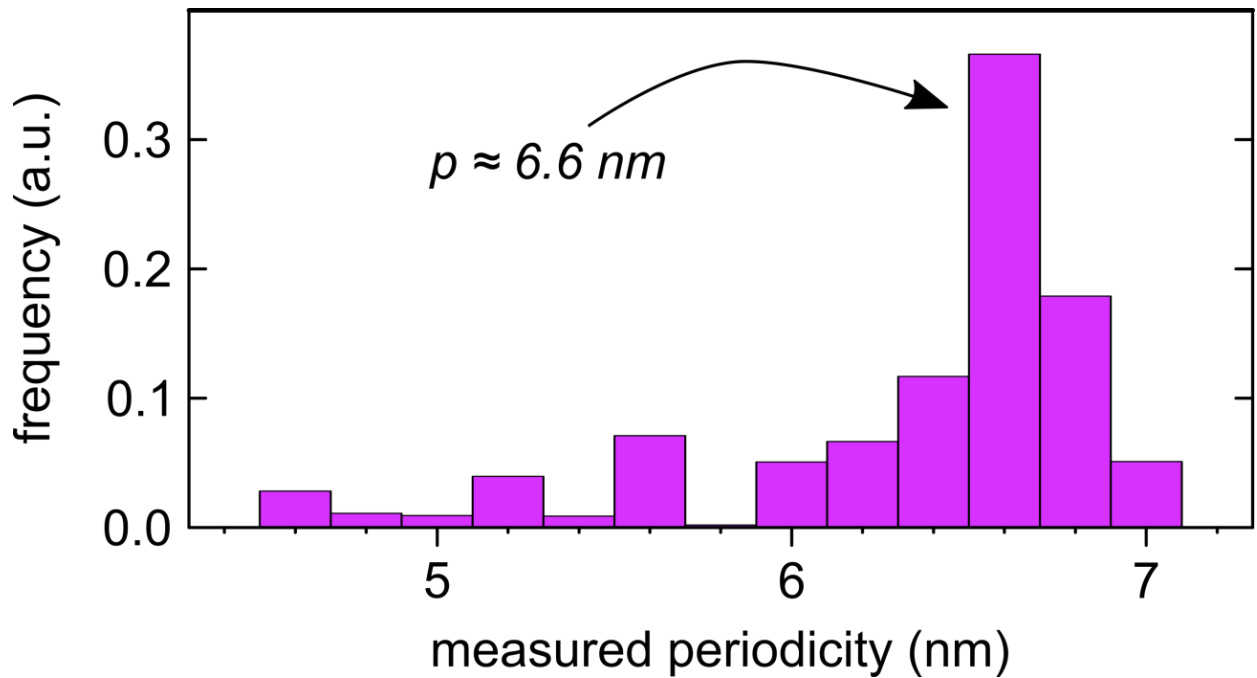

Supplementary Fig. S5: A histogram of the normalized frequency of observations topographical modulations in an FFTEM experiment on CB6OBO6CB, with the measured periodicity weighted by the domain area from various locations throughout the sample. The histogram exhibits a peak at $p_{tri,FFTEM}$ = 6.6 nm, demonstrating that this is most likely to be the low-temperature helical pitch exhibited by the TB phase.

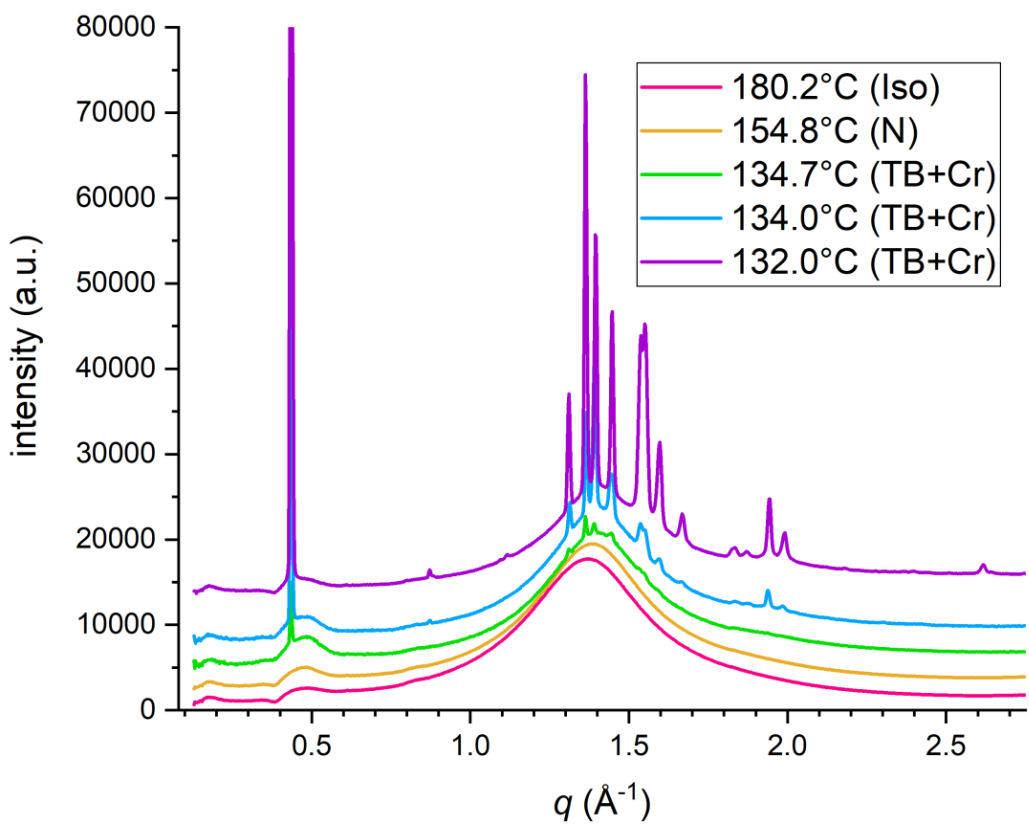

Supplementary Fig. S6: 1D-WAXS scans of CB6OBO6CB on cooling from the isotropic phase. In the isotropic phase, CB6OBO6CB exhibits diffuse features at $q_{z,\text{WAXS}} = 0.48$ Å$^{-1}$ and $q_{\perp,\text{WAXS}} = 1.37$ Å$^{-1}$. The $q_{z,\text{WAXS}}$ peak sharpens slightly in the nematic phase, as expected. On further cooling, we capture the TB-Cr coexistence at 134.7°C, with the $q_{z,\text{WAXS}}$ feature sharpening up from the nematic phase, but not shifting significantly in $q$. This feature indicates the presence of a modest density modulation, or pseudo-layering, along the helical direction, with a period of $s_{\text{tri}} = 13$ Å. In this experiment, the sample crystallizes very soon after the formation of the monotropic TB phase, with the TB phase and the crystal phase coexisting in the lowest temperature scans. In the crystal phase, a sharp scattering peak appears near the $q_{z,\text{WAXS}}$ feature, indicating lamellar crystalline ordering.

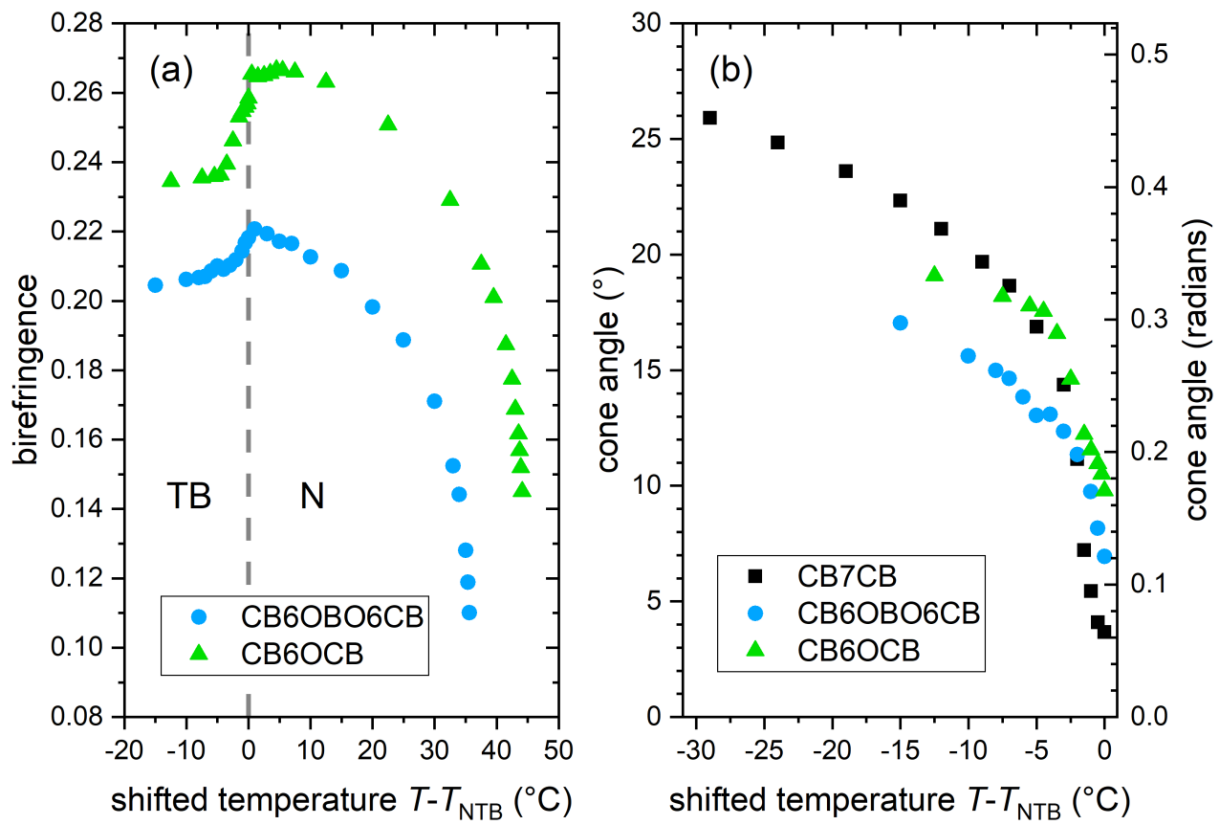

Supplementary Fig. S7: Birefringence measurements and the calculated optical cone angle of CB6OCB and CB6OBO6CB. (a) We measured the birefringence of the dimer and trimer using a Berek optical compensator. We then used the procedure developed by Meyer *et al.*(23) to calculate the optically-derived cone angle $\theta_{optical}$ (b) from the birefringence. For comparison, we include our cone angle measurement of CB7CB first reported in ref.(19), and consistent with other measurements(23, 33).

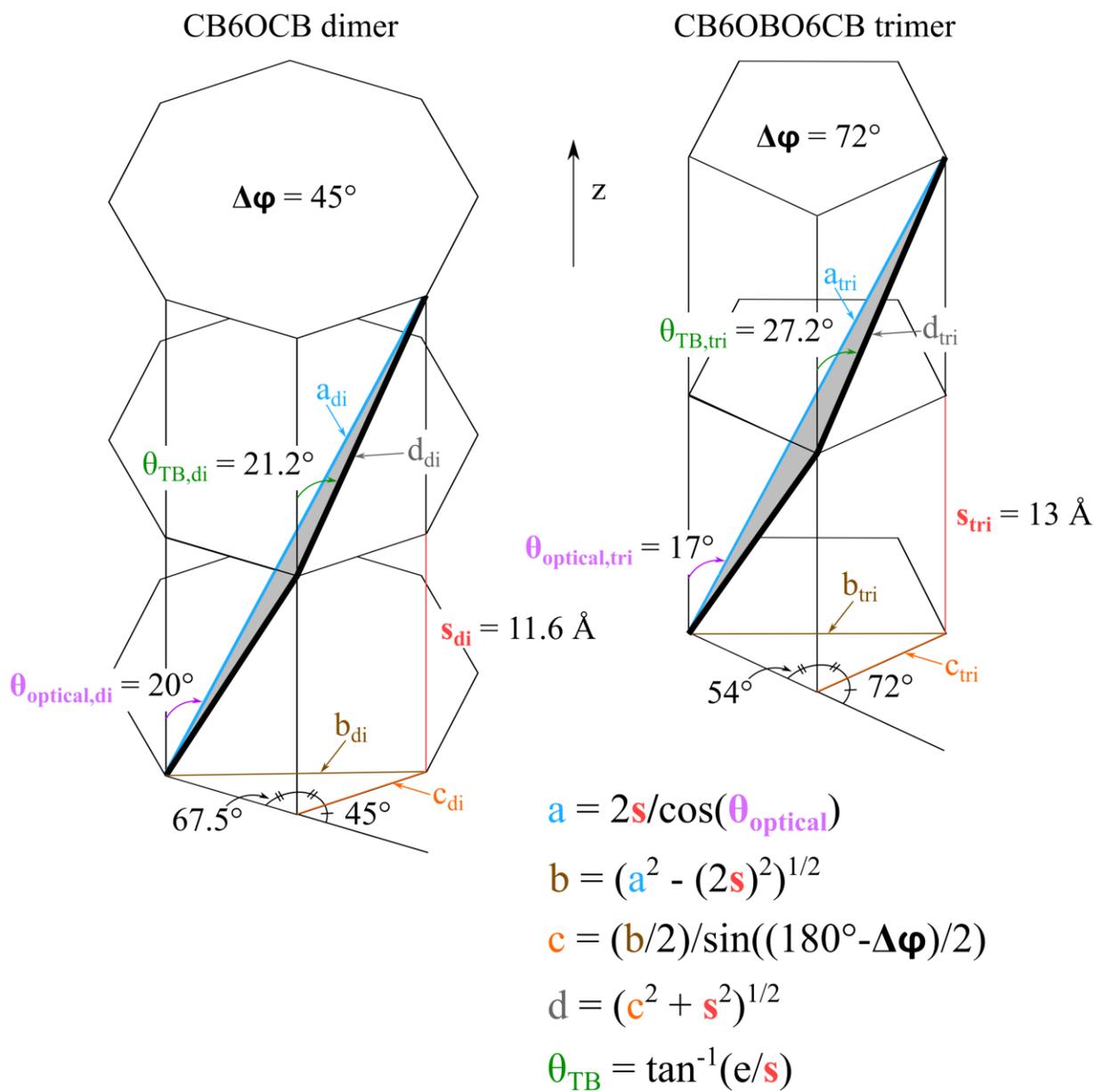

Supplementary Fig. S8: Derivation of the true TB cone angle $\theta_{TB}$ from the at low temperature experimentally obtained geometric parameters: $\theta_{optical}$, $s$, and $\Delta\varphi$ (in bold in the schematic). $\theta_{optical}$ measures the tilt of the molecular plane away from z, whereas $\theta_{TB}$ measures the tilt of the local director (the direction that a single monomer segment of the molecule points) away from z. These schematics are not necessarily to scale.

37